\documentclass[aps,prl,twocolumn,groupedaddress,superscriptaddress,showpacs]{revtex4}

\usepackage{graphicx}
\usepackage{color}
\usepackage{amsmath}
\usepackage{amsfonts}
\usepackage{amssymb}
\usepackage{dcolumn}
\usepackage{float}

\begin{document}
  \title{Explosive percolation via control of the largest cluster}

  \author{N. A. M. Ara\'ujo}
    \email{nuno@ethz.ch}
    \affiliation{Computational Physics for Engineering Materials, IfB, ETH Zurich, Schafmattstr. 6, 8093 Zurich, Switzerland}

  \author{H. J. Herrmann}
    \email{hans@ifb.baug.ethz.ch}
    \affiliation{Computational Physics for Engineering Materials, IfB, ETH Zurich, Schafmattstr. 6, 8093 Zurich, Switzerland}
    \affiliation{Departamento de F\'isica, Universidade Federal do Cear\'a, Campus do Pici, 60451-970 Fortaleza, Cear\'a, Brazil}


  \begin{abstract}
    We show that only considering the largest cluster suffices to obtain a first-order percolation transition.
    As opposed to previous realizations of explosive percolation our models obtain Gaussian cluster distributions and compact clusters as one would expect at first-order transitions.
    We also discover that the cluster perimeters are fractal at the transition point, yielding a fractal dimension of $1.23\pm0.03$, close to that of watersheds.
  \end{abstract}

  \maketitle

  Percolation, the paradigm for random connectivity, has since Hammersley \cite{Broadbent57} been one of the most often applied statistical models \cite{Stauffer94,Sahimi94}.
  Its phase transition being related to magnetic models \cite{Fortuin72} is in all dimensions one of the most robust second-order transitions known.
  This explains the enormous excitement generated by the recent work by Achlioptas, D'Souza, and Spencer \cite{Achlioptas09} describing a stochastic rule apparently yielding a discontinuous percolation transition on a fully connected graph.
  Subsequent work applied the process on other networks \cite{Ziff09,Ziff10,Radicchi09,Radicchi10,Cho09,Cho10}.
  However, reported results of finite-size studies and size distributions are not consistent with a first-order transition.
  Since then various rules have been devised \cite{Friedman10,Manna10,DSouza10} and even a Hamiltonian formalism was proposed \cite{Moreira10}, all attempting a discontinuous transition towards an infinite cluster.
  In all proposed models one tries to keep the clusters of similar size and some authors additionally suppress the internal bonds of clusters \cite{Achlioptas09,Moreira10}.
  Could one obtain a clear and consistent first-order percolation transition?
  It is the objective of the present Letter to answer this question.
  One criterion is the cluster size distribution at the percolation threshold.
  Radicchi and Fortunato \cite{Radicchi10} as well as Ziff \cite{Ziff10} found a power-law distribution with an exponent close to two.
  Although, different from the exponent of classical percolation the sole fact of finding a power law is untypical for first-order transitions.
  Also unusual for a first-order transition is that the clusters are fractal, as we found happens for the Achlioptas rule, from the behavior of the order parameter with the system size \cite{Ziff10,Radicchi10}.
  It is a purpose of the present Letter to present a model in which a Gaussian cluster size distribution and compact clusters can be achieved in a systematic way, characterized by a fractal perimeter yielding a fractal dimension similar to the one of watersheds and random polymers in strongly disordered media.

    \begin{figure}
      \begin{tabular}{cc}
        \includegraphics[width=0.24\textwidth]{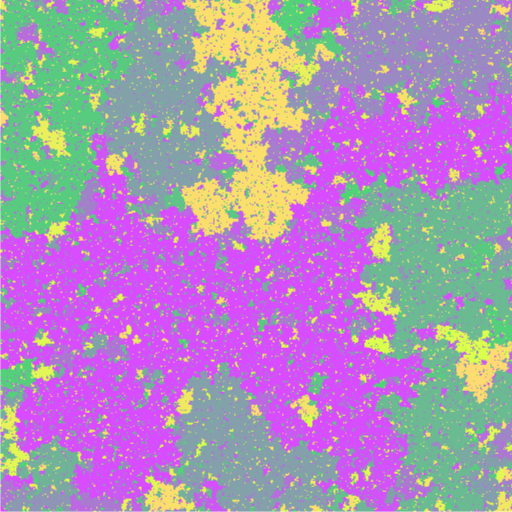} &
        \includegraphics[width=0.24\textwidth]{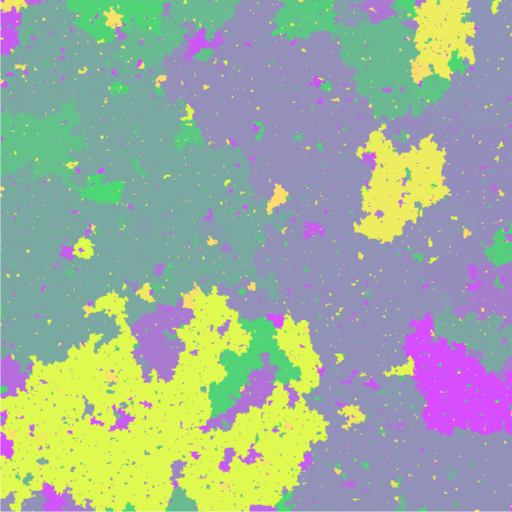} \\
        {\it classical} & {\it product rule} \\ 
        \includegraphics[width=0.24\textwidth]{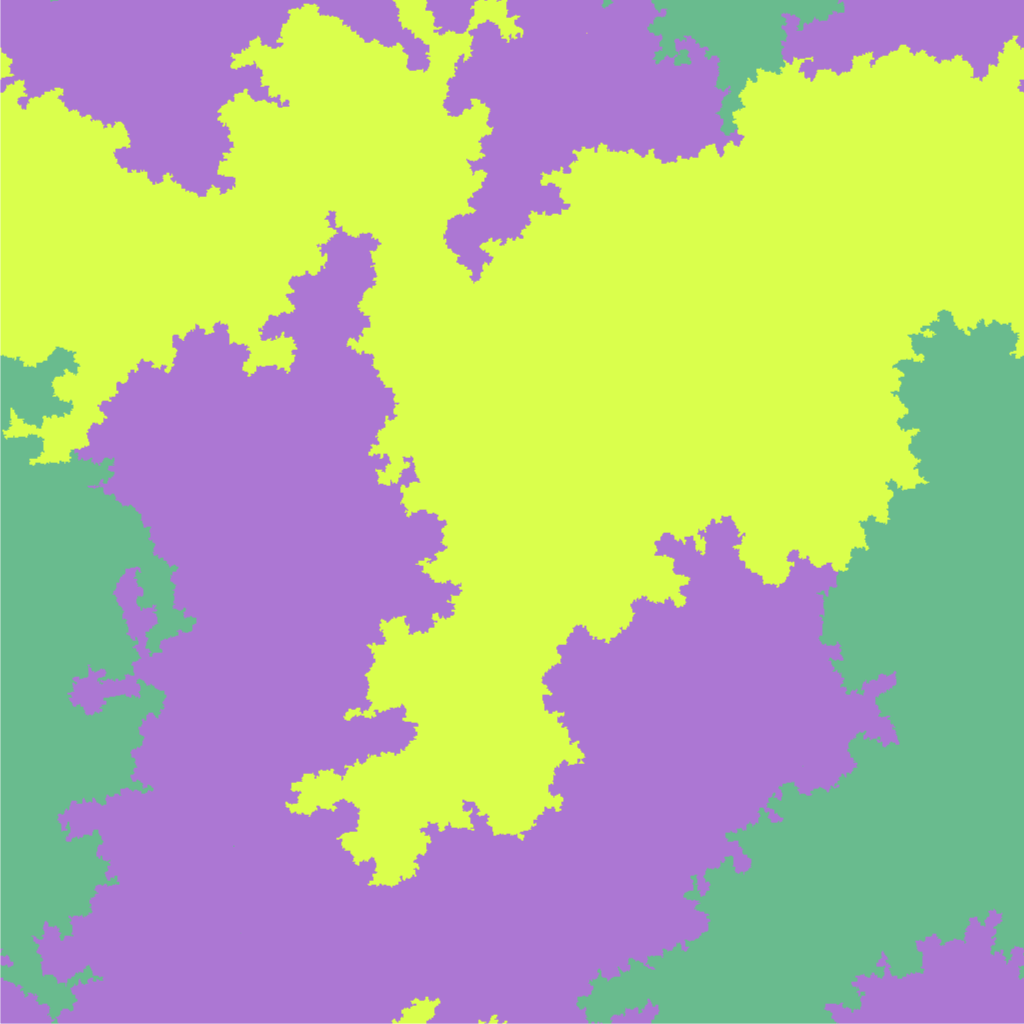} &
        \includegraphics[width=0.24\textwidth]{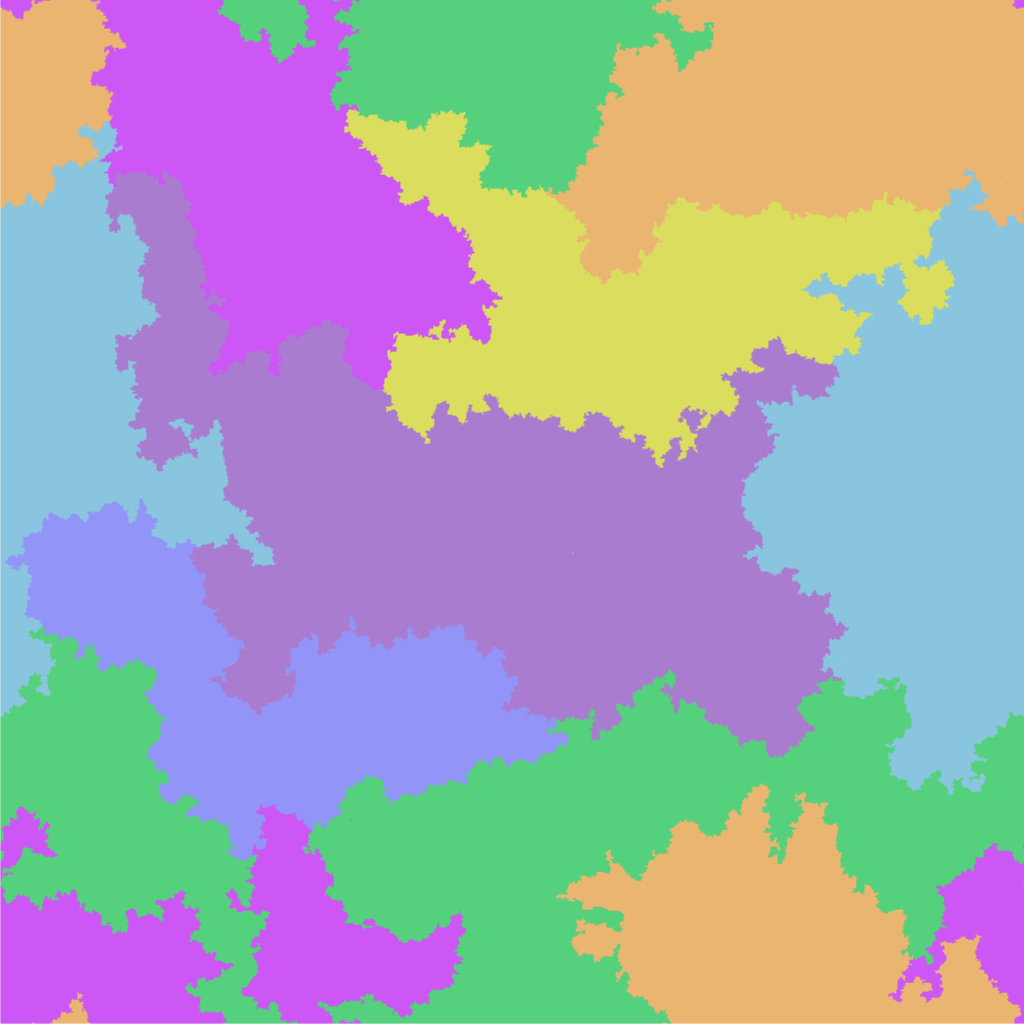} \\
        {\it largest cluster} & {\it Gaussian} \\
      \end{tabular}
      \caption{ (Color online) Snapshots of the system, obtained on a square lattice with $1024^2$ sites, at $p_c$, for four different bond percolation models, namely, classical \cite{Stauffer94}, Achlioptas product rule \cite{Ziff09}, {\it largest cluster} model ($\alpha= 1$), and {\it Gaussian} model ($\alpha=1$).
                The {\it largest cluster} and {\it Gaussian} models are introduced in this Letter.
                \label{fig::snap}}
    \end{figure}

  Usual bond percolation can be implemented on a square lattice by randomly occupying bonds between neighboring sites, reaching its threshold at a certain fraction when opposite borders are first connected through one large cluster \cite{Stauffer79,Isichenko92,Stauffer94}.
  This percolation threshold, is characterized by the continuous vanishing of the order parameter, i.e., a second-order transition.
  On a fully connected graph, Achlioptas {\it et al.} \cite{Achlioptas09}, used the {\it best-of-two} product rule studied in detail by Friedman and Landsberg \cite{Friedman10}. 
  Ziff reported simulations on a regular square lattice \cite{Ziff09,Ziff10}, while Radicchi and Fortunato \cite{Radicchi09,Radicchi10} and Cho {\it et al.} \cite{Cho09} on scale-free networks.

  More recently, other approaches have been introduced to obtain explosive percolation.
  Instead of a {\it best-of-two} rule Manna {\it et al.} \cite{Manna10}, Cho {\it et al.} \cite{Cho10}, and Moreira {\it et al.} \cite{Moreira10} proposed a {\it weighted} rule where bonds are occupied according to a certain probability.
  However, despite being {\it rejection-free} schemes, they are limited to small-system sizes and/or reduced number of samples.
  Here, we suggest an acceptance method where new bonds are selected randomly and occupied according to a certain weight yielding, for the first time, a clear first-order transition.
  The considered scheme allows to consider system sizes $64$ times larger than before \cite{Manna10}, specifically, we consider systems of $4096^2$ sites and averages over $10^4$ samples.

  In our simplest rule (``{\it largest cluster} model''), as for classical bond percolation, a link is randomly selected among the empty ones.
  If its occupation would not lead to the formation or growth of the largest cluster, it is always occupied, otherwise, it is occupied with probability

    \begin{equation}\label{eq::acceptance.probability}
      \mbox{min}\left\{1, \exp\left[-\alpha \left(\frac{s-\bar{s}}{\bar{s}}\right)^2\right]\right\} \ \ ,
    \end{equation}
 
\noindent where $s$ is the size of the cluster that would be formed by occupying this bond and $\bar{s}$ the average cluster size after occupying the bond.
  The parameter $\alpha$ controls the allowed size dispersion.
  Note that, for $\alpha\leq 0$, since the size of the largest cluster is always greater (or equal) than the average cluster size, all new bonds are occupied reducing to classical bond percolation, characterized by a continuous transition at the percolation threshold \cite{Stauffer94}.
  For $\alpha>0$, the probability of Eq.~(\ref{eq::acceptance.probability}) suppresses the formation of a cluster significantly larger than the average, inducing a homogenization of cluster sizes.
  The Gaussian function has been considered because this is what we expect for the cluster size distribution at a first-order transition.
  However, to observe a discontinuous transition any other function could be chosen, as far as it constrains the largest cluster differing significantly, in size, from the average cluster.

    \begin{figure}
      \includegraphics[width=\columnwidth]{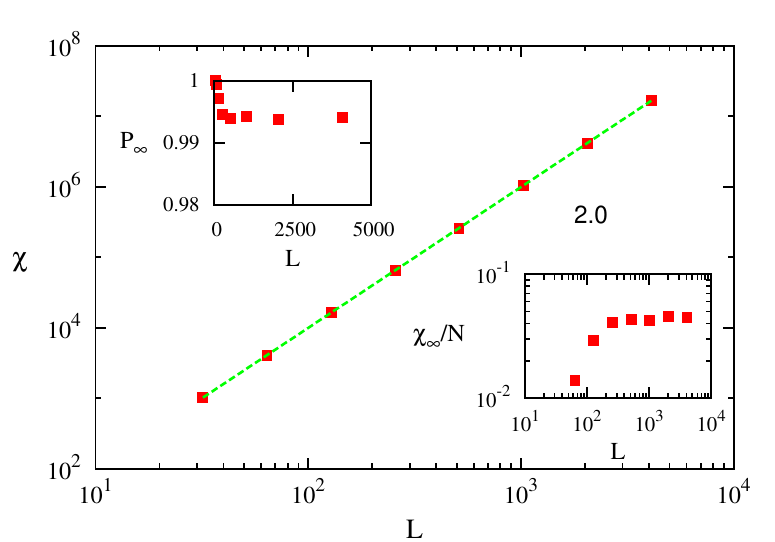}
      \caption{
        (Color online) Size dependence, for the {\it largest cluster} model, of the susceptibility ($\chi$), fraction of sites in the largest cluster ($P_\infty$), and its standard deviation per site ($\chi_\infty/N$) at the percolation threshold, on a square lattice of linear size ($L$) ranging from $32$ to $4096$.
        All bonds are occupied with the same probability except the ones that lead to the formation/growth of the largest cluster, to which an occupation probability $q$ is assigned, Eq.~(\ref{eq::acceptance.probability}), with $\alpha=1$.
        Results have been averaged over $10^4$ samples.
        \label{fig::avoidlarg}}
    \end{figure}

  For nonequilibrium problems, where a free energy cannot be defined, transitions can still be classified based on the behavior of the order parameter \cite{Odor04}.
  A first-order transition, is characterized by a jump in the order parameter, otherwise, a transition is denoted as continuous.
  For percolation, we define as order parameter the fraction of sites in the largest cluster ($P_\infty$) \cite{Stauffer94}.
  Here we also consider two other quantities: the second moment of the cluster size distribution ($\chi$), defined as

    \begin{equation}\label{eq::sec.mom}
      \chi= \sum_{i} s^2_i \ \ ,
    \end{equation}

  \noindent where the sum runs over all clusters $i$, and the standard deviation ($\chi_\infty$) of the largest cluster size ($s_{max}$) over different samples,
 
    \begin{equation}\label{eq::susc.inf.clus}
      \chi_\infty= \sqrt{\langle s^2_{max}\rangle-\langle s_{max}\rangle^2} \ \ .
    \end{equation}

  \noindent To estimate the percolation threshold we consider the average value of $p$ (fraction of occupied bonds) at which a connected path linking opposite boundaries of the system is obtained.
  Considering different system sizes, for $\alpha=1$, we obtain for the percolation threshold $p_c=0.632\pm0.002$.
  To identify the order of the transition, in the {\it largest cluster} model, Fig.~\ref{fig::avoidlarg} presents a finite-size study for $P_\infty$, $\chi$, and $\chi_\infty/N$, averaged over $10^4$ samples of square lattices with linear sizes ranging from $32$ to $4096$.
  As we can see in the top inset of Fig.~\ref{fig::avoidlarg}, above a certain system size, the order parameter, at the percolation threshold, does not show any finite-size dependence, staying at a constant value in the thermodynamic limit ($L\rightarrow\infty$).
  The second moment of the cluster size distribution ($\chi$) scales with $L^d (d=2)$ which is a sign of a first-order transition \cite{Binder81,Binder84}.
  The standard deviation of the largest cluster ($s_{max}$) per lattice site, which was also considered in Refs.~\cite{Ziff09} and \cite{Ziff10}, converges, for larger system sizes, to a constant value, corroborating the presence of a discontinuous transition.

      \begin{figure}
        \includegraphics[width=\columnwidth]{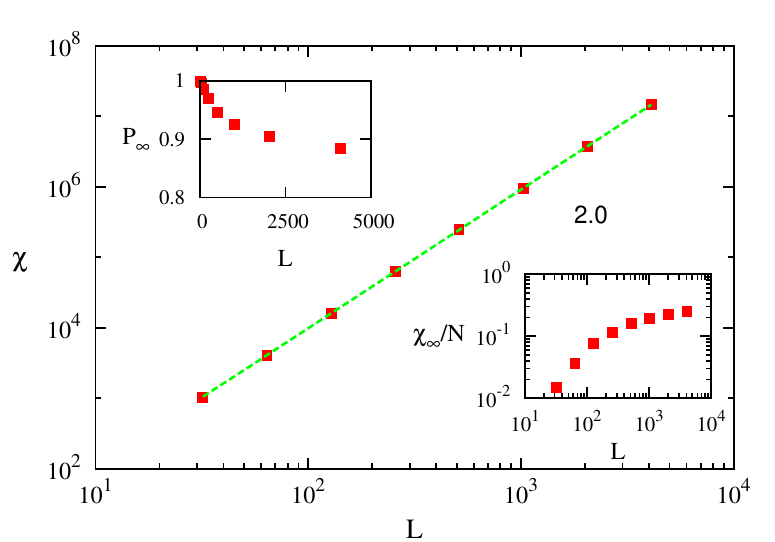}
          \caption{
            (Color online) Size dependence, for the {\it Gaussian} model, with $\alpha=1$, of the susceptibility ($\chi$), fraction of sites in the largest cluster ($P_\infty$), and its standard deviation per site ($\chi_\infty/N$) at the percolation threshold, on a square lattice of linear size ($L$) ranging from $32$ to $4096$.
            All bonds are occupied with a probability given by Eq.~(\ref{eq::acceptance.probability}).
            Results have been averaged over $10^4$ samples.
          \label{fig::gaussian}}
      \end{figure}

  To explicitly control the cluster size distribution we also implemented the following model.
  A new bond is chosen from the list of empty ones and occupied with probability given by Eq.~(\ref{eq::acceptance.probability}).
  For internal connections we consider $s$ as twice the cluster size.
  Since equation~(\ref{eq::acceptance.probability}) is a Gaussian with average size $\bar{s}$ and size dispersion $\bar{s}/\sqrt{2\alpha}$, we denote this model as {\it Gaussian} model.
  Note that here the occupation probability is assigned to all new bonds even when they are not related to the largest cluster.
  This not only guarantees the control over clusters greater than the average, as in the previous model, but also over the smaller ones.
  For $\alpha= 0$, all bonds have the same probability and, therefore, the model reduces to classical bond percolation.
  For negative $\alpha$, the growth of larger clusters is favored in two different ways: they differ more from the average value and have more empty bonds than the smaller ones.
  Yet, for all negative $\alpha$, the model recovers the classical universality class of percolation \cite{Stauffer94,Odor04}.

  As example, for positive $\alpha$, we present, in Fig.~\ref{fig::gaussian}, a size dependence study of the order parameter, second moment of the cluster size distribution, and standard deviation per site of the largest cluster, for the {\it Gaussian} model, with $\alpha=1$, at the percolation threshold, on a regular square lattice with linear size ($L$) ranging from $32$ to $4096$.
  Results were averaged over $10^4$ samples.
  We extrapolate, for the infinite system, a percolation threshold $p_c=0.56244\pm0.00006$.
  As for the {\it largest cluster} model, the density of the infinite cluster does not change significantly with the system size, the second moment of the cluster size distribution scales with $L^d (d=2)$, and the standard deviation per site of the largest cluster converges to a non-zero constant.
  As before these results imply a first-order transition.

  Figure~\ref{fig::snap} shows snapshots for four different models of bond percolation: classical, product rule, {\it largest cluster} model, and {\it Gaussian} model.
  All figures have been obtained at their respective percolation thresholds ($p_c$).
  For classical percolation and for the product rule, clusters of very different sizes are obtained.
  In fact, the cluster size distribution is characterized by a power law \cite{Radicchi10,Ziff10}.
  However, for the {\it largest cluster} and the {\it Gaussian} model, a characteristic cluster size is observed.
  Both models lead to a localized cluster size distribution.
  Small size dispersion and number of clusters are observed for the {\it largest cluster} model.
  According to Eq.~(\ref{eq::acceptance.probability}), increasing the value of $\alpha$ decreases the size dispersion.

\begin{figure}
      \includegraphics[width=\columnwidth]{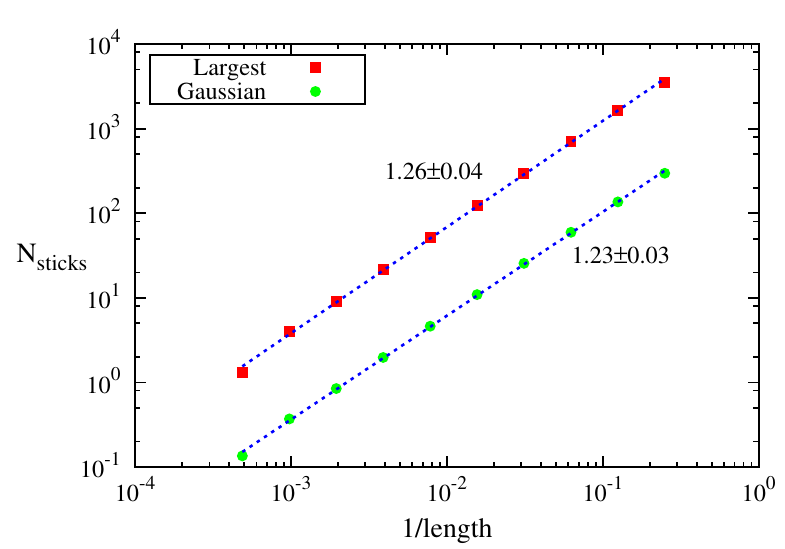}
      \caption{
                (Color online) Number of sticks necessary to follow the perimeter of the infinite cluster as a function of the stick length, to obtain the fractal dimension of the perimeter with the {\it yardstick method}.
                For both the {\it largest cluster} and {\it Gaussian} models, with $\alpha= 1$.
                For the {\it Gaussian} model data were vertically shifted by a factor of $0.1$.
                Results have been averaged over $10^4$ samples of lattices with linear size $2048$.
        \label{fig::fracdim}}
    \end{figure}

  As clearly seen in the snapshots of Fig.~\ref{fig::snap}, clusters obtained with our models are compact but we find that the surface is fractal.
  For the {\it Gaussian} model, we calculate for the cluster perimeter a fractal dimension of $1.23\pm0.03$, obtained with the {\it yardstick method} \cite{Tricot88} (Fig.~\ref{fig::fracdim}).
  For the {\it largest cluster} model, it is also characterized by a fractal perimeter with a fractal dimension of $1.26\pm0.04$ (Fig.~\ref{fig::fracdim}).
  Compact clusters with fractal surface were also reported for irreversible aggregation growth in the limit of high concentration by Kolb {\it et al.} \cite{Kolb87}.
  For the present models, the percolation thresholds are larger than the ones from previous models due to the compactness of the clusters.

\begin{figure}
      \includegraphics[width=\columnwidth]{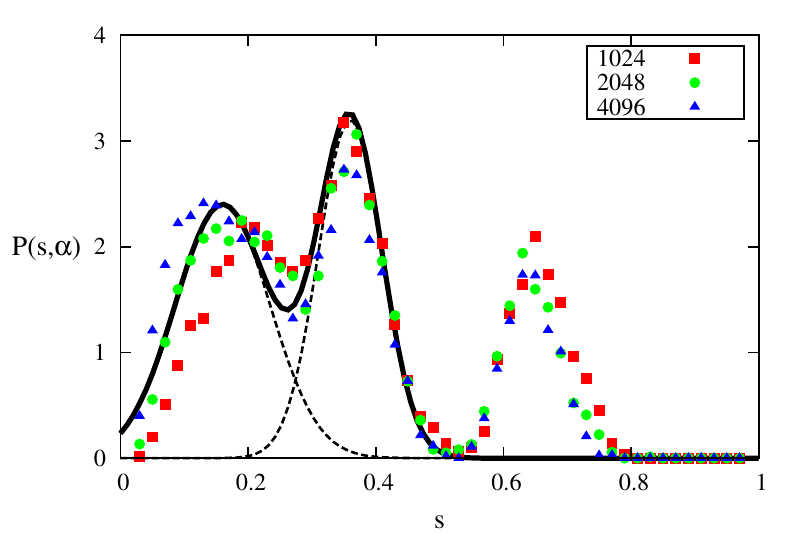}
      \caption{ (Color online) Cluster size distribution for the {\it Gaussian} model for different system sizes ($\alpha=1$), at the percolation threshold, on a square lattice, averaged over $10^4$ samples.
                Black-dashed lines are two Gaussian distributions fitting the results from simulation.
                The black-solid line is the sum of both curves.
        \label{fig::sizedist}}
    \end{figure}

  In Fig.~\ref{fig::sizedist} we see the cluster size distribution, $P(s,\alpha)$, for different system sizes, obtained with the {\it Gaussian} model.
  Measurements have been performed at the percolation threshold on a square lattice with $1024^2$, $2048^2$, and $4096^2$ sites, and averaged over $10^4$ samples.
  Three characteristic peaks are observed.
  In fact, the third peak (around $0.7$) is only due to the largest cluster and only appears due to the small number of clusters at the percolation threshold, being finite-size effect.
  This peak is not observed when we compute the same distribution without considering the largest cluster.
  In the thermodynamic limit, since an infinite number of clusters exists, the contribution of a single cluster to the distribution vanishes.
  The presence of two main peaks is characteristic for a first-order transition showing, for a finite system, at the percolation threshold, coexistence of the percolative and non-percolative states \cite{Binder92}.

  In conclusion, the present work reveals that, to obtain explosive percolation on a regular lattice it is sufficient to control the formation and growth of the largest cluster, instead of applying a rule to the overall set of empty bonds.
  We propose the {\it largest cluster} model which systematically suppresses the formation of a largest cluster.
  We introduce as well, the {\it Gaussian} model, where a weight is assigned to each selected bond, such that a Gaussian distribution of cluster sizes is obtained, revealing the coexistence of two states at the percolation threshold.
  Our models, yielding clear first-order transitions, show that explosive percolation can be obtained under less stringent conditions that previously thought shedding light on the minimum ingredients to trigger explosive percolation. 
  In fact, we believe that our restrictions on the formation of a largest cluster differing significantly, in size, from the average, is the required necessary condition and hope that this statement can one day be formally proven. 
  The value of the novel fractal dimension of percolation that we discovered in the cluster perimeters is intriguingly close to the one found for watersheds ($1.211\pm0.001$) \cite{Fehr09} and random polymers in strongly disordered media ($1.22\pm0.02$) \cite{Porto99}.
  
\begin{acknowledgments}
We acknowledge financial support from the ETH Competence Center Coping with Crises in Complex Socio-Economic Systems (CCSS) through ETH Research Grant CH1-01-08-2.
\end{acknowledgments}

\bibliography{text}

\end{document}